\documentclass[letterpaper]{article} 
\nonstopmode
\usepackage{aaai25}  
\usepackage{mathptmx}
\usepackage{helvet}  
\usepackage{courier}  
\usepackage[hyphens]{url}  
\usepackage{graphicx} 
\usepackage{natbib}  
\usepackage{caption} 
\usepackage{algorithm}
\usepackage{algorithmic}
\usepackage{adjustbox}
\usepackage{array}
\usepackage{longtable}
\usepackage{newfloat}
\usepackage{listings}
\DeclareCaptionStyle{ruled}{labelfont=normalfont,labelsep=colon,strut=off} 
\floatstyle{ruled}
\newfloat{listing}{tb}{lst}{}
\floatname{listing}{Listing}
\title{Applications and Advances of Artificial Intelligence in Music Generation:A Review}

\author{
    Yanxu Chen,\textsuperscript{\rm 1}
    Linshu Huang,\textsuperscript{\rm 2}
    Tian Gou\textsuperscript{\rm 3}
}
\affiliations{
    \textsuperscript{\rm 1}Keystone Academy, Beijing, China\\
    \textsuperscript{\rm 2}Xi{-}an Jiaotong{-}Liverpool University, Suzhou, China\\
    \textsuperscript{\rm 3}XrayBot AI Lab, Beijing, China\\
    yanxu.chen@student.keystoneacademy.cn, linshu.huang22@student.xjtlu.edu.cn, goutian@xraybot.com
}

\usepackage{bibentry}
\nocopyright
\begin{document}

\maketitle

\begin{abstract}
In recent years, artificial intelligence (AI) has made significant progress in the field of music generation, driving innovation in music creation and applications. This paper provides a systematic review of the latest research advancements in AI music generation, covering key technologies, models, datasets, evaluation methods, and their practical applications across various fields. The main contributions of this review include: (1) presenting a comprehensive summary framework that systematically categorizes and compares different technological approaches, including symbolic generation, audio generation, and hybrid models, helping readers better understand the full spectrum of technologies in the field; (2) offering an extensive survey of current literature, covering emerging topics such as multimodal datasets and emotion expression evaluation, providing a broad reference for related research; (3) conducting a detailed analysis of the practical impact of AI music generation in various application domains, particularly in real-time interaction and interdisciplinary applications, offering new perspectives and insights; (4) summarizing the existing challenges and limitations of music quality evaluation methods and proposing potential future research directions, aiming to promote the standardization and broader adoption of evaluation techniques. Through these innovative summaries and analyses, this paper serves as a comprehensive reference tool for researchers and practitioners in AI music generation, while also outlining future directions for the field.
\end{abstract}

%

\section{Introduction}

Music, as a universal and profound art form, transcends cultural and geographical boundaries, playing an unparalleled role in emotional expression\citep{juslin2011handbook}. With the rapid advancement of technology, music creation has evolved from the manual operations of the early 20th century, relying on analog devices and tape recordings, to today’s fully digital production environment\citep{katz2010capturing,pinch2012oxford,deruty2022development,oliver2022digital}. In this evolution, the introduction of Artificial Intelligence (AI) has injected new vitality into music creation, driving the rapid development of automatic music generation technologies and bringing unprecedented opportunities for innovation\citep{briot2020deep,zhang2023artificial}.

\textbf{Research Background and Current Status:}The research on automatic music generation dates back more than 60 years, with the earliest attempts primarily based on grammatical rules and probabilistic models\citep{hiller1979experimental,dash2023ai}. However, with the rise of deep learning technologies, the field of AI music generation has entered an unprecedented period of prosperity\citep{goodfellow2016deep,moysis2023music}. Modern AI technologies can not only handle symbolic music data but also generate high-fidelity audio content directly, with applications ranging from traditional instrument simulation to entirely new sound design\citep{oord2016wavenet,lei2024comprehensive}. Symbolic music generation relies on representations such as piano rolls and MIDI, enabling the creation of complex structured musical compositions; meanwhile, audio generation models deal directly with continuous audio signals, producing realistic and layered sounds\citep{dong2018musegan,ji2023survey}.

In recent years, AI music generation technologies have made remarkable progress, especially in the areas of model architecture and generation quality\citep{huang2018music,agostinelli2023musiclm}. The application of Generative Adversarial Networks (GANs), Transformer architectures, and the latest diffusion models has provided strong support for the diversity, structure, and expressiveness of generated music\citep{goodfellow2014generative,vaswani2017attention,ho2020denoising,kong2020diffwave,shahriar2022gan}. Additionally, new hybrid model frameworks that combine the strengths of symbolic and audio generation further enhance the structural integrity and timbral expressiveness of generated music\citep{huang2018music,wang2024whole,qian2024musicaog}. These advancements have not only expanded the technical boundaries of AI music generation but also opened up new possibilities for music creation\citep{wang2024review}.

\textbf{Research Motivation:}Despite significant advances in AI music generation, numerous challenges remain. Enhancing the originality and diversity of generated music, capturing long-term dependencies and complex structures in music, and developing more standardized evaluation methods are core issues that the field urgently needs to address. Furthermore, as the application areas of AI-generated music continue to expand—such as healthcare, content creation, and education—the demands for quality and control of generated music are also increasing. These challenges provide a broad space for future research and technological innovation.

\begin{figure}
    \centering
    \includegraphics[width=1\linewidth]{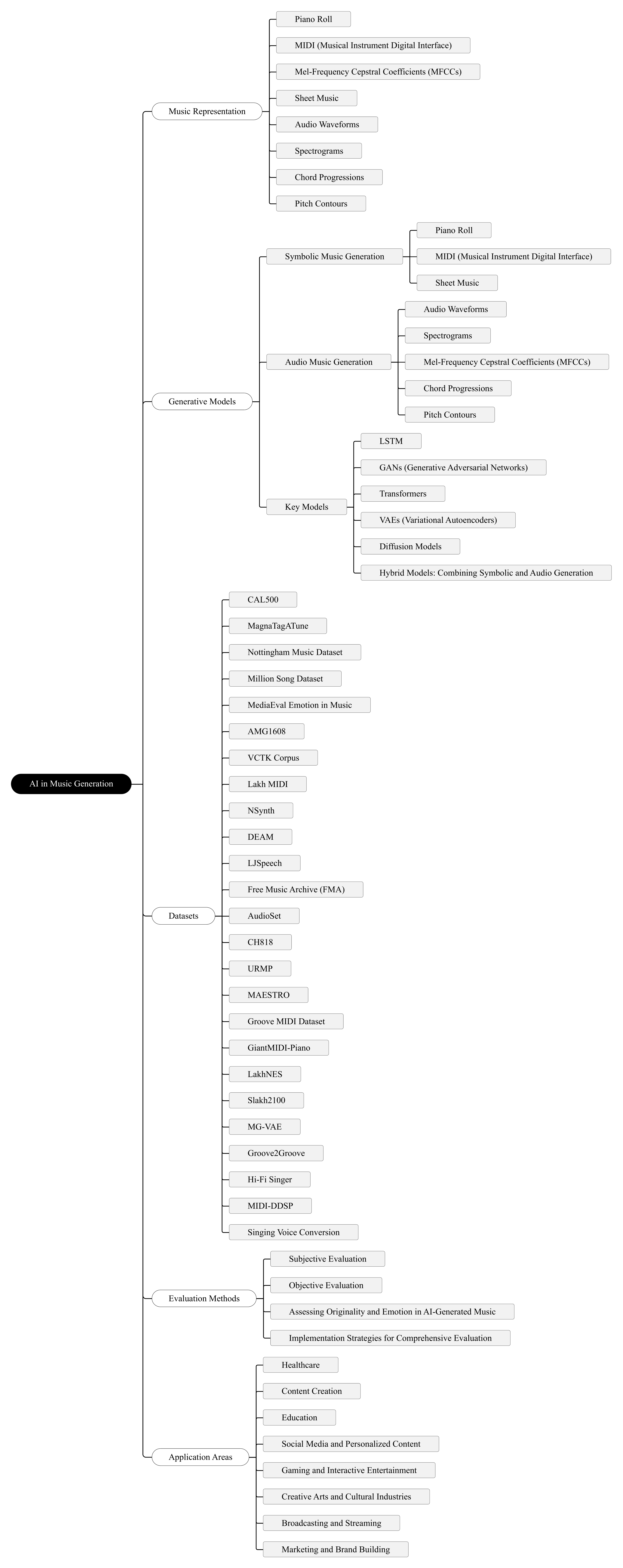}
    \caption{The core logic of our review}
    \label{fig:1}
\end{figure}

\textbf{Research Objectives:}This paper aims to systematically review the latest research progress in symbolic and audio music generation, explore their potential and challenges in various application scenarios, and forecast future development directions. Through a comprehensive analysis of existing technologies and methods, this paper seeks to provide valuable references for researchers and practitioners in the AI music generation field and inspire further innovation and exploration. We hope that this research will promote the continuous innovation of AI in music creation, making it a core tool in music production in the future.The core logic of this review paper is illustrated in Figure \ref{fig:1}.

\section{History of Music Production}

\textbf{Early Stages of Music Production}

In the early 20th century, music production mainly relied on analog equipment and tape recording technology. Sound engineers and producers used large analog consoles for recording, mixing, and mastering. This period emphasized the craftsmanship and artistry of live performances, with the constraints of recording technology and equipment making the process of capturing each note filled with uncertainty and randomness.\citep{zak2001poetics,schmidtHorning2013chasing}
The introduction of synthesizers brought revolutionary changes to music creation, particularly in electronic music. In the 1970s, synthesizers became increasingly popular, with brands like Moog and Roland symbolizing the era of electronic music. Synthesizers generated various sounds by modulating waveforms (such as sine and triangle waves), allowing music producers to create a wide range of tones and effects on a single instrument, thereby greatly expanding the possibilities for musical expression\citep{PinchTrocco+2004,holmes2012electronic}.

\textbf{The Rise of Digital Audio Workstations (DAWs)}

With advances in digital technology, Digital Audio Workstations (DAWs) began to rise in the late 1980s and early 1990s. The advent of DAWs marked the transition of music production into the digital era, integrating recording, mixing, editing, and composition into a single software platform, making the music production process more efficient and convenient\citep{hracs2016production,danielsen2018music,theberge2021sound,cross2023music}.
The widespread application of MIDI (Musical Instrument Digital Interface) further propelled the development of digital music production. MIDI facilitated communication between digital instruments and computers, becoming a critical tool in modern music production. Renowned DAWs like Logic Pro, Ableton Live, and FL Studio provided producers with integrated working environments, streamlining the music creation process and democratizing music production\citep{derrico2016interface,reuter2022daws}.

\textbf{Expansion of Plugins and Virtual Instruments}

The popularity of DAWs fueled the development of plugins and virtual instruments. Plugins, as software extensions, added new functionalities or sound effects to DAWs, vastly expanding the creative potential of music production. Platforms like Kontakt offered various high-quality virtual instruments, while synthesizer plugins such as Serum and Phase Plant, utilizing advanced wavetable synthesis, provided producers with extensive sound design possibilities. The diversity and flexibility of plugins greatly broadened the creative space of music production, enabling producers to modulate, edit, and layer various sound effects within a single software environment\cite{tanev2013virtual,wang2017design,rambarran2021virtual}.

\textbf{Application of Artificial Intelligence in Music Production}

With technological advancement, Artificial Intelligence (AI) has gradually entered the field of music production. AI technologies can analyze large volumes of music data, extract patterns and features, and generate new music compositions. Max/MSP, an early interactive audio programming environment, allowed users to create their own sound effects and instruments through coding, marking the initial application of AI technology in music production\citep{tan2021tutorial,hernandezolivan2022music,ford2024reflection,marschall2007machine,privato2022creative}.

As AI technologies matured, machine learning-based tools emerged, capable of generating music based on given datasets and automating tasks such as mixing and mastering. Modern AI music generation technologies can not only simulate existing styles but also create entirely new musical forms, opening up new possibilities for music creation\citep{taylor2024exploration}.

\textbf{Trends in Modern Music Production}

Today’s music production is fully digital, with producers able to complete every step from composition to mastering within a DAW. The diversity and complexity of plugins continue to grow, including vocoders, resonators, and convolution reverbs, bringing infinite possibilities to music creation. The introduction of AI has further pushed the boundaries of music creation, making automation and intelligent production a reality\citep{briot2020deep,agostinelli2023musiclm}.
Modern music production is not only the result of technological accumulation but also a model of the fusion of art and technology. The incorporation of AI technologies has enriched the music creation toolbox and spurred the emergence of new musical styles, making music creation more diverse and dynamic\citep{deruty2022development,tao2022new,goswami2023music}.

\section{Music Representation}
The representation of music data is a core component of AI music generation systems, directly influencing the quality and diversity of the generated results. Different music representation methods capture distinct characteristics of music, significantly affecting the input and output of AI models. Below are some commonly used music representation methods and their application scenarios:

\textbf{3.1 Piano Roll}

A piano roll is a two-dimensional matrix that visually represents the notes and timing of music, making it particularly suitable for capturing melody and chord structures. The rows of the matrix represent pitch, columns represent time, and the values indicate whether a particular pitch is activated at a given time point. This representation is widely used in deep learning models as it directly maps to the input and output layers of neural networks, facilitating the processing and generation of complex musical structures. For example, MuseGAN\citep{dong2018musegan} uses piano roll representation for multi-part music, generating harmonically rich compositions through Generative Adversarial Networks (GANs).

\textbf{3.2 MIDI (Musical Instrument Digital Interface)}

MIDI is a digital protocol used to describe various musical parameters such as notes, pitch, velocity, tempo, and chords. MIDI files do not record actual audio data but rather instructions that control audio, making them highly flexible and allowing playback in various styles on different synthesizers and virtual instruments. MIDI is extensively used in music creation, arrangement, and AI music generation, especially in symbolic music generation, where it serves as a crucial format for input and output data. Its advantages lie in cross-platform and cross-device compatibility and the precise control of musical parameters. MusicVAE\citep{brunner2018midivaemodelingdynamicsinstrumentation} utilizes MIDI to represent symbolic music, where notes and timing are discrete, enabling the model to better capture structural features and generate music with complex harmony and melody.

\textbf{3.3 Mel Frequency Cepstral Coefficients (MFCCs)}

MFCCs are a compact representation of the spectral characteristics of audio signals, widely used in speech and music processing, particularly effective in capturing subtle differences in music. By decomposing audio signals into short-time frames and applying the Mel frequency scale, MFCCs capture audio features perceived by the human ear. Although primarily used in speech recognition, MFCCs also find extensive applications in music emotion analysis, style classification, and audio signal processing. For example, Google’s NSynth project uses MFCCs\citep{engel2017neural} for generating and classifying different timbres.

\textbf{3.4 Sheet Music}

Sheet music is a traditional form of music representation that records musical information through staff notation and various musical symbols. It includes not only pitch and rhythm but also dynamics, expressive marks, and other performance instructions. In AI music generation, sheet music representation is also employed, particularly for generating readable compositions that adhere to music theory. Models using sheet music as input, such as Music Transformer\citep{huang2018musictransformer}, can generate compositions with complex structure and coherence.

\textbf{3.5 Audio Waveform}

The audio waveform directly represents the time-domain waveform of audio signals, suitable for generating and processing actual audio data. Although waveform representation involves large data volumes and complex processing, it provides the most raw and detailed audio information, crucial in audio synthesis and sound design. For instance, the WaveNet\citep{oord2016wavenetgenerativemodelraw} model uses waveforms directly to generate highly realistic speech and music.

\textbf{3.6 Spectrogram}

A spectrogram converts audio signals into a frequency domain representation, showing how the spectrum of frequencies evolves over time. Common spectrograms include Short-Time Fourier Transform (STFT) spectrograms, Mel spectrograms, and Constant-Q transform spectrograms. Spectrograms are highly useful in music analysis, classification, and generation, as they capture both the frequency structure and temporal characteristics of audio signals. The Tacotron 2\citep{wang2017tacotron} model uses spectrograms as intermediate representations for generating audio from text, transforming text input into Mel spectrograms and then using WaveNet to generate the final waveform audio. The DDSP model\citep{engel2020ddsp} employs spectrograms as intermediate representations to generate high-quality audio by manipulating frequency domain signals. It combines traditional Digital Signal Processing (DSP) techniques with deep learning models to generate realistic instrument timbres and complex audio effects, making it highly effective in music generation and sound design.

\textbf{3.7 Chord Progressions}

Chord progressions are sequences of chords that represent changes over time and are crucial in popular, jazz, and classical music. AI music generation systems can learn patterns of chord progressions to generate harmonious and structured music. For example, the ChordGAN model\citep{lu2021chordgan} generates chord progressions for background harmonies in popular music.

\textbf{3.8 Pitch Contour}

Pitch contour represents the variation of pitch over time, particularly useful for analyzing and generating melodic lines. Pitch contours capture subtle pitch changes in music, aiding in generating smooth and natural melodies. OpenAI’s Jukebox model\citep{dhariwal2020jukebox} uses pitch contours to generate complete songs with coordinated melodies and background accompaniment.

\section{Generative Models}

The field of AI music generation can be divided into two main directions: symbolic music generation and audio music generation. These two approaches correspond to different levels and forms of music creation.
\begin{figure*}[htbp] 
    \centering
    \includegraphics[width=0.68\textwidth]{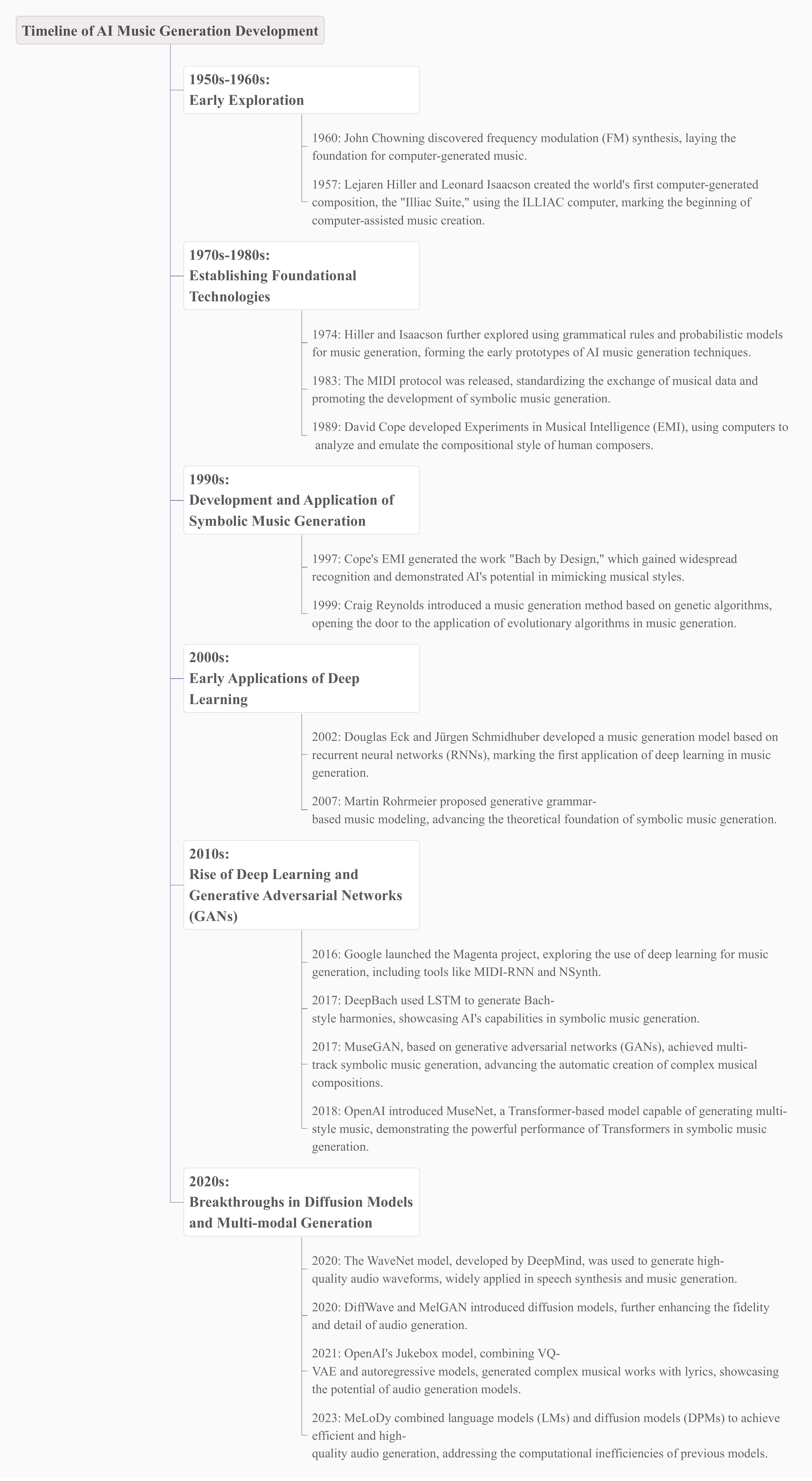}
    \caption{Timeline of AI Music Generation Development}
    \label{fig:2}
\end{figure*}

\textbf{4.1 Symbolic Music Generation}

Symbolic music generation uses AI technologies to create symbolic representations of music, such as MIDI files, sheet music, or piano rolls. The core of this approach lies in learning the structures of music, chord progressions, melodies, and rhythmic patterns to generate compositions with logical and structured music. These models typically handle discrete note data, and the generated results can be directly played or further converted into audio.
In symbolic music generation, LSTM models have shown strong capabilities. For instance, DeepBach\citep{pmlr-v70-hadjeres17a} uses LSTMs to generate Bach-style harmonies, producing harmonious chord progressions based on given musical fragments. However, symbolic music generation faces challenges in capturing long-term dependencies and complex structures, particularly when generating music on the scale of entire movements or songs, where maintaining long-range musical dependencies can be difficult.

Recently, Transformer-based symbolic music generation models have demonstrated more efficient capabilities in capturing long-term dependencies. For example, the Pop Music Transformer\citep{huang2020pop} combines self-attention mechanisms and Transformer architecture to achieve significant improvements in generating pop music. Additionally, MuseGAN, a GAN-based multi-track symbolic music generation system, can generate multi-part music suitable for creating compositions with rich layers and complex harmonies. The MuseCoco model\citep{lu2023musecoco} combines natural language processing with music creation, generating symbolic music from text descriptions and allowing precise control over musical elements, making it ideal for creating complex symbolic music works. However, symbolic music generation mainly focuses on notes and structure, with limited control over timbre and expressiveness, highlighting its limitations.

\textbf{4.2 Audio Music Generation}

Audio music generation directly generates the audio signal of music, including waveforms and spectrograms, handling continuous audio signals that can be played back directly or used for audio processing. This approach is closer to the recording and mixing stages in music production, capable of producing music content with complex timbres and realism.

WaveNet\citep{oord2016wavenetgenerativemodelraw}, a deep learning-based generative model, captures subtle variations in audio signals to generate expressive music audio, widely used in speech synthesis and music generation. Jukebox\citep{dhariwal2020jukebox}, developed by OpenAI, combines VQ-VAE and autoregressive models to generate complete songs with lyrics and complex structures, with sound quality and expressiveness approaching real recordings. However, audio music generation typically requires substantial computational resources, especially when handling large amounts of audio data. Additionally, audio generation models face challenges in controlling the structure and logic of music over extended durations.

Recent research on diffusion models has made significant progress, initially used for image generation but now extended to audio. For example, DiffWave\citep{kong2020diffwave} and WaveGrad\citep{chen2020wavegrad} are two representative audio generation models; the former generates high-fidelity audio through a progressive denoising process, and the latter produces detailed audio through a similar diffusion process. The MeLoDy model\citep{stefani1987melody} combines language models (LMs) and diffusion probability models (DPMs), reducing the number of forward passes while maintaining high audio quality, addressing computational efficiency issues. Noise2Music\citep{huang2023noise2music}, based on diffusion models, focuses on the correlation between text prompts and generated music, demonstrating the ability to generate music closely related to input text descriptions.

Overall, symbolic music generation and audio music generation represent the two primary directions of AI music generation. Symbolic music generation is suited for handling and generating structured, interpretable music, while audio music generation focuses more on the details and expressiveness of audio signals. Future research could combine these two methods to enhance the expressiveness and practicality of AI music generation, achieving seamless transitions from symbolic to audio, and providing more comprehensive technical support for music creation.

\textbf{4.3 Current Major Types of Generative Models}

The core of AI music generation lies in using different generative models to simulate and create music. Each model has its unique strengths and application scenarios. Below are some major generative models and their applications:

\textbf{Long Short-Term Memory Networks (LSTM): }LSTM excels in handling sequential data with temporal dependencies, effectively capturing long-term dependencies in music and generating coherent and expressive music sequences. Models like BachBot\citep{liang2016bachbot} and DeepBach\citep{hadjeres2017deepbach} utilize LSTMs to generate Bach-style music, demonstrating LSTM's strong capabilities in music generation. However, LSTM models often require large amounts of data for training and have relatively high computational costs, limiting their application in resource-constrained environments.

\textbf{Generative Adversarial Networks (GAN):} GANs generate high-quality, realistic music content through adversarial training between a generator and a discriminator, making them particularly suitable for generating complex and diverse audio. For instance, DCGAN\citep{radford2016unsupervisedrepresentationlearningdeep} excels in generating high-fidelity audio. Models like WaveGAN\citep{donahue2019adversarialaudiosynthesis} and MuseGAN\citep{ji2023survey} have made significant progress in single-part and multi-part music generation, respectively. MusicGen\citep{copet2024simplecontrollablemusicgeneration}, developed by Meta, is a deep learning-based music generation model capable of producing high-quality, diverse music fragments from noise or specific input conditions. However, GANs can have unstable training processes and may suffer from mode collapse, leading to a lack of diversity in the generated music.

\textbf{Transformer Architecture: }Transformers leverage self-attention mechanisms to efficiently process sequential data, particularly adept at capturing long-range dependencies and complex structures in music compositions. Notable work includes the Music Transformer\citep{huang2018music}, which uses self-attention to generate structured music segments, effectively capturing motifs and repetitive structures across multiple time scales. This results in music that is structurally coherent and closer to human compositional styles. MusicLM\citep{agostinelli2023musiclm} combines Transformer-based language models with audio generation, offering innovation in generating high-fidelity music audio from text descriptions. However, Transformer models require substantial computational resources for training and generation.

\textbf{Variational Autoencoders (VAE): }VAEs generate new data points by learning latent representations, suitable for tasks involving diversity and creativity in music generation. The MIDI-VAE model\citep{brunner2018midivaemodelingdynamicsinstrumentation} uses VAE for music style transfer, demonstrating the potential of VAE in generating diverse music. The Conditional VAE (CVAE) enhances diversity by introducing conditional information, reducing mode collapse risks. OpenAI's Jukebox\citep{dhariwal2020jukebox} combines Vector Quantized VAE (VQ-VAE-2) with autoregressive models to generate complete songs with lyrics and complex structures. Compared to GANs or Transformers, VAE-generated music may lack musicality and coherence.

\textbf{Diffusion Models: }Diffusion models generate high-quality audio content by gradually removing noise, making them suitable for high-fidelity music generation. Recent research includes the Riffusion model\citep{Forsgren_Martiros_2022}, utilizing the Stable Diffusion model for real-time music generation, producing music in various styles from text prompts or image conditions; Moûsai\citep{schneider-etal-2024-mousai}, a diffusion-based music generation system, generates persistent, high-quality music from text prompts in real time. The lengthy training and generation processes of diffusion models can limit their application in real-time music generation scenarios.

\textbf{Other Models and Methods:} Besides the models mentioned above, Convolutional Neural Networks (CNNs), other types of Recurrent Neural Networks (RNNs), and methods combining multiple models have also been applied in music generation. Additionally, rule-based methods and evolutionary algorithms offer diverse technical and creative approaches for music generation. For example, WaveNet\citep{oord2016wavenet}, a CNN-based model, is innovative in directly modeling audio signals. MelGAN\citep{kumar2019melgan} uses efficient convolutional architectures to generate detailed audio.

\textbf{4.4 Hybrid Model Framework:Integrating Symbolic and Audio Music Generation}

Recently, researchers have recognized that combining the strengths of symbolic and audio music generation can significantly enhance the overall quality of generated music. Symbolic music generation models (e.g., MIDI or sheet music generation models) excel at capturing musical structure and logic, while audio generation models (e.g., WaveNet\citep{oord2016wavenet} or Jukebox\citep{dhariwal2020jukebox}) focus on generating high-fidelity and complex timbre audio signals. However, each model type has distinct limitations: symbolic generation models often lack expressiveness in timbre, and audio generation models struggle with long-range structural modeling.
To address these challenges, recent studies have proposed hybrid model frameworks that combine the advantages of symbolic and audio generation. A common strategy is to use methods that jointly employ Variational Autoencoders (VAE) and Transformers. For example, in models like MuseNet\citep{topirceanu2014musenet} and MusicVAE\citep{yang2019inspecting}, symbolic music is first generated by a Transformer and then converted into audio signals. These models typically use VAE to capture latent representations of music and employ Transformers to generate sequential symbolic representations.
Self-supervised learning methods have gained increasing attention in symbolic music generation. These approaches often involve pre-training models to capture structural information in music, which are then applied to downstream tasks. Models like Jukebox\citep{dhariwal2020jukebox} use self-supervised learning to enhance the generalization and robustness of generative models.

Additionally, combining hierarchical symbolic music generation with cascaded diffusion models has proven effective\citep{wang2024whole}. This approach defines a hierarchical music language to capture semantic and contextual dependencies at different levels. The high-level language handles the overall structure of a song, such as paragraphs and phrases, while the low-level language focuses on notes, chords, and local patterns. Cascaded diffusion models train at each level, with each layer’s output conditioned on the preceding layer, enabling control over both the global structure and local details of the generated music.

The fusion of symbolic and audio generation frameworks combines symbolic representations with audio signals, resulting in music that is not only structurally coherent but also rich in timbre and detailed expression. The symbolic generation part ensures harmony and logic, while the audio generation part adds complex timbre and dynamic changes, paving the way for creating high-quality and multi-layered music.Examples of related work for different foundational models are shown in Table \ref{table:1}. The development trajectory of AI music generation technology can be seen in Figure \ref{fig:2}.
\begin{table*}[ht]
\centering
\begin{adjustbox}{max width=\textwidth}
\begin{tabular}{|m{2.5cm}|m{3cm}|m{4cm}|m{4cm}|m{4cm}|}
\hline
\textbf{Model Type} & \textbf{Related Research} & \textbf{Strengths} & \textbf{Challenges} & \textbf{Suitable Scenarios}\\ 
\hline
LSTM & DeepBach,
BachBot& Good at capturing temporal dependencies and sequential data& High computational cost, training requires large datasets, struggles with long-term dependencies & Suitable for sequential music generation tasks, such as harmonization and melody generation \\ 
\hline
GAN & MuseGAN,
WaveGAN& High-quality, realistic generation, suitable for complex and diverse audio & Training can be unstable, prone to mode collapse, limited in capturing structure and long-term dependencies & Ideal for generating complex audio content like multi-instrument music or diverse sound effects \\ 
\hline
Transformer & Music Transformer,
MusicLM& Excellent at capturing long-range dependencies and complex structures & High computational demand, requires large amounts of data for training & Best for generating music with complex structures, long sequences, and coherent compositions \\ 
\hline
VAE & MIDI-VAE,
Jukebox& Encourages diversity and creativity, suitable for style transfer & Generated music can lack musical coherence and expressiveness compared to GANs or Transformers & Best for tasks requiring high variability and creativity, such as style transfer and music exploration \\ 
\hline
Diffusion Models & DiffWave,WaveGrad,
Noise2Music& High-quality audio generation, excels in producing high-fidelity music & Training and generation time can be long, challenging in real-time scenarios & Suitable for generating high-quality audio and sound effects, particularly in media production \\ 
\hline
Hybrid Models & MuseNet,
MusicVAE& Combines strengths of symbolic and audio models, controls structure and timbre & Complexity in integrating different model types, requires more sophisticated tuning & Ideal for creating music that requires both structural coherence and rich audio expressiveness, useful in advanced music composition \\ 
\hline
\end{tabular}
\end{adjustbox}
\caption{Comparison of Various Generative Models for Music Generation}
\label{table:1}

\end{table*}

\begin{table*}[ht]
\centering
\begin{adjustbox}{max width=\textwidth}
\begin{tabular}{|p{2.8cm}|p{2.8cm}|p{3.5cm}|p{3cm}|p{3cm}|p{1cm}|}
\hline
\textbf{Model Name} & \textbf{Base Architecture} & \textbf{Dataset Used} & \textbf{Data Representation} & \textbf{Loss Function} & \textbf{Year} \\
\hline
WaveNet & CNN & VCTK Corpus, YouTube Data & Waveform & L1 Loss & 2016 \\
\hline
BachBot & LSTM & Bach Chorale Dataset & Symbolic Data & Cross-Entropy Loss & 2016 \\
\hline
DCGAN & CNN & Lakh MIDI Dataset (LMD) & Audio Waveform & Binary Cross-Entropy Loss & 2016 \\
\hline
DeepBach & LSTM & Bach Chorale Dataset & MIDI File & Cross-Entropy Loss & 2017 \\
\hline
MuseGAN & GAN & Lakh MIDI Dataset (LMD) & Multi-track MIDI & Binary Cross-Entropy Loss & 2018 \\
\hline
MIDI-VAE & VAE & MIDI files (Classic, Jazz, Pop, Bach, Mozart) & Pitch roll, Velocity roll, Instrument roll & Cross Entropy, MSE, KL Divergence & 2018 \\
\hline
Music Transformer & Transformer & Lakh MIDI Dataset (LMD) & MIDI File & Cross-Entropy Loss & 2019 \\
\hline
WaveGAN & GAN & Speech Commands, AudioSet & Audio Waveform & GAN Loss (Wasserstein Distance) & 2019 \\
\hline
Jukebox & VQ-VAE + Autoregressive & 1.2 million songs (LyricWiki) & Audio Waveform & Reconstruction Loss, Perceptual Loss & 2019 \\
\hline
MelGAN & GAN-based & VCTK, LJSpeech & Audio Waveform & GAN Loss (Multi-Scale Discriminator) & 2019 \\
\hline
Pop Music Transformer & Transformer-XL & Custom Dataset (Pop piano music) & REMI (Rhythm-Event-Metric Information) & Cross-Entropy Loss & 2020 \\
\hline
DiffWave & Diffusion Model & VCTK, LJSpeech & Waveform & L1 loss, GAN loss & 2020 \\
\hline
Riffusion & Diffusion + CLIP & Large-Scale Popular Music Dataset (Custom) & Spectrogram Image & Diffusion Loss, Reconstruction Loss & 2022 \\
\hline
MusicLM & Transformer + AudioLDM & Free Music Archive (FMA) & Audio Waveform & Cross-Entropy Loss, Contrastive Loss & 2023 \\
\hline
MusicGen & Transformer & Shutterstock, Pond5 & Audio Waveform & Cross-Entropy Loss, Perceptual Loss & 2023 \\
\hline
Music ControlNet & Diffusion Model & MusicCaps (~1800 hours) & Audio Waveform & Diffusion Loss & 2023 \\
\hline
Moûsai & Diffusion Model & Moûsai-2023 & Mel-spectrogram & Spectral loss, GAN loss & 2023 \\
\hline
MeLoDy & LM-guided Diffusion & 257k hours of non-vocal music & Audio Waveform & Cross-Entropy Loss, Diffusion Loss & 2023 \\
\hline
MuseCoco & GAN-based & Multiple MIDI datasets including Lakh MIDI and MetaMIDI & Multi-track MIDI & Binary Cross-Entropy Loss & 2023 \\
\hline
Noise2Music & Diffusion Model & MusicCaps, MTAT, Audioset & Audio Waveform & Diffusion Loss & 2023 \\
\hline
\end{tabular}
\end{adjustbox}
\caption{Representative Music Generation Models: Key Features and Technical Details}
\label{table:2}
\end{table*}

\section{Datasets}
In the field of AI music generation, the choice and use of datasets profoundly impact model performance and the quality of generated results. Datasets not only provide the foundation for model training but also play a key role in enhancing the diversity, style, and expressiveness of generated music. This section introduces commonly used datasets in AI music generation and discusses their characteristics and application scenarios.

\textbf{5.1 Commonly Used Open-Source Datasets for Music Generation}

In the music generation domain, the following datasets are widely used resources that cover various research directions, from emotion recognition to audio synthesis. This section introduces these datasets, including their developers or owners, and briefly describes their specific applications.

\textbf{•	CAL500 (2007)}

The CAL500 dataset\citep{turnbull2007towards}, developed by Gert Lanckriet and his team at the University of California, San Diego, contains 500 MP3 songs, each with detailed emotion tags. These tags are collected through subjective evaluations by listeners, covering various emotional categories. The dataset is highly valuable for static emotion recognition and emotion analysis research.

\textbf{•	MagnaTagATune (MTAT) (2008)}

Developed by Edith Law, Kris West, Michael Mandel, Mert Bay, and J. Stephen Downie, the MagnaTagATune dataset\citep{law2009evaluation} uses an online game called "TagATune" to collect data. It contains approximately 25,863 audio clips, each 29 seconds long, sourced from Magnatune.com songs. Each clip is associated with a binary vector of 188 tags, independently annotated by multiple players. This dataset is widely used in automatic music annotation, emotion recognition, and instrument classification research.

\textbf{•	Nottingham Music Dataset (2009)}

The Nottingham Music Dataset\citep{boulanger2012modeling} was originally developed by Eric Foxley at the University of Nottingham and released on SourceForge. It includes over 1,000 traditional folk tunes suitable for ABC notation. The dataset has been widely used in traditional music generation, music style analysis, and symbolic music research.

\textbf{•	Million Song Dataset (MSD) (2011)}

The Million Song Dataset\citep{bertin2011million} is a benchmark dataset designed for large-scale music information retrieval research, providing a wealth of processed music features without including original audio or lyrics. It is commonly used in music recommendation systems and feature extraction algorithms.

\textbf{•	MediaEval Emotion in Music (2013)}

The MediaEval Emotion in Music dataset\citep{soleymani20131000} contains 1,000 MP3 songs specifically for music emotion recognition research. The emotion tags were obtained through subjective evaluations by a group of annotators, making it useful for developing and validating music emotion recognition models.

\textbf{•	AMG1608 (2015)}

The AMG1608 dataset\citep{Penha2015AMG1608}, developed by Carmen Penha, Fabio G. Cozman, and researchers from the University of São Paulo, contains 1,608 music clips, each 30 seconds long, annotated for emotions by 665 subjects. The dataset is particularly suitable for personalized music emotion recognition research due to its detailed emotional annotations, especially those provided by 46 subjects who annotated over 150 songs.

\textbf{•	VCTK Corpus (2016)}

Developed by the CSTR laboratory at the University of Edinburgh, the VCTK Corpus\citep{VCTK2017} contains speech data recorded by 110 native English speakers with different accents. Each speaker read about 400 sentences, including texts from news articles, rainbow passages, and accent archives. This dataset is widely used in Automatic Speech Recognition (ASR) and Text-to-Speech (TTS) model development.

\textbf{•	Lakh MIDI (2017)}

The Lakh MIDI dataset\citep{raffel2016learning} is a collection of 176,581 unique MIDI files, with 45,129 files matched and aligned with entries from the Million Song Dataset. It is designed to facilitate large-scale music information retrieval, including symbolic (using MIDI files only) and audio-based (using information extracted from MIDI files as annotations for matching audio files) research.

\textbf{•	NSynth (2017)}

NSynth\citep{engel2017neural}, developed by Google’s Magenta team, is a large-scale audio dataset containing over 300,000 monophonic sound samples generated using instruments from commercial sample libraries. Each note has unique pitch, timbre, and envelope characteristics, sampled at 16 kHz and lasting 4 seconds. The dataset includes notes from various instruments sampled at different pitches and velocities.

\textbf{•	DEAM (2017)}

The DEAM dataset\citep{aljanaki2017developing}, developed by a research team at the University of Geneva, is specifically designed for dynamic emotion recognition in music. It contains 1,802 musical pieces, including 1,744 45-second music clips and 58 full songs, covering genres such as rock, pop, electronic, country, and jazz. The songs are annotated with dynamic valence and arousal values over time, providing insights into the dynamic changes in musical emotion.

\textbf{•	LJSpeech (2017)}

The LJSpeech dataset\citep{ljspeech17} is a public domain speech dataset consisting of 13,100 short audio clips of a single speaker reading from seven non-fiction books. Each clip has a corresponding transcription, with lengths ranging from 1 to 10 seconds and totaling about 24 hours. The texts were published between 1884 and 1964 and are in the public domain.

\textbf{•	Free Music Archive (FMA) (2017)}

FMA\citep{defferrard2017fmadatasetmusicanalysis}, developed by Michaël Defferrard and others from Ecole Polytechnique Fédérale de Lausanne (EPFL), is a large-scale music dataset sourced from the Free Music Archive (FMA). It contains 106,574 music tracks spanning 161 different genres, with high-quality full-length audio, rich metadata, precomputed audio features, and hierarchical genre labels. FMA is widely used in music classification, retrieval, style recognition, and audio feature extraction research.

\textbf{•	AudioSet (2017)}

AudioSet\citep{gemmeke2017audio}, developed by Google, is a large-scale audio dataset containing over 2 million labeled 10-second audio clips collected from YouTube videos. The dataset uses a hierarchical ontology of 635 audio categories, covering various everyday sound events. Due to its broad audio categories and high-quality annotations, AudioSet is an important benchmark for audio event detection, classification, and multimodal learning.

\textbf{•	CH818 (2017)}

The CH818 dataset\citep{hu2017cross} contains 818 Chinese pop music clips annotated with emotion labels, mainly used for emotion-driven music generation and pop music style analysis. Despite challenges in annotation consistency, the dataset offers valuable resources for music generation and emotion recognition research in Chinese contexts.

\textbf{•	URMP  Dataset (2018)}

The URMP dataset\citep{li2018creating} is designed to facilitate audio-visual analysis of music performance. It includes 44 multi-instrument music pieces composed of individually recorded tracks synchronized for ensemble performance. The dataset provides MIDI scores, high-quality individual instrument recordings, and ensemble performance videos.

\textbf{•	MAESTRO (2018)}

MAESTRO (MIDI and Audio Edited for Synchronous Tracks and Organization)\citep{hawthorne2018enabling} is a dataset developed by Google AI, containing over 200 hours of aligned MIDI and audio recordings primarily sourced from international piano competitions. The MIDI data includes details like velocity and pedal controls, precisely aligned (~3 ms) with high-quality audio (44.1–48 kHz 16-bit PCM stereo), making it an essential resource for music generation and automatic piano transcription research.

\textbf{•	Groove MIDI Dataset (GMD) (2019)}

The Groove MIDI Dataset\citep{gillick2019learning} contains 13.6 hours of MIDI and audio data recording human-performed drum performances. Recorded with a Roland TD-11 V-Drum electronic drum kit, it includes 1,150 MIDI files and over 22,000 measures of drum grooves played by 10 drummers, including professionals and amateurs.

\textbf{•	GiantMIDI-Piano (2020)}

The GiantMIDI-Piano dataset\citep{kong2020giantmidi} comprises 10,855 solo piano pieces' MIDI files, automatically transcribed from real recordings using a high-resolution piano transcription system. The dataset covers a rich repertoire from 2,786 composers and accurately captures musical details like pitch, onset, offset, and dynamics, making it a valuable resource for piano music generation, transcription, and music analysis.

\textbf{•	LakhNES (2019)}

Developed by Chris Donahue, the LakhNES dataset\citep{donahue2019lakhnes} is a large MIDI dataset focused on pre-training language models for multi-instrument music generation. It combines data from the Lakh MIDI and NES-MDB datasets, providing diverse and unique training material suitable for complex Transformer architectures in cross-domain multi-instrument music generation tasks.

\textbf{•	Slakh2100 (2019)}

The Slakh2100 dataset\citep{manilow2019cutting} consists of MIDI compositions and synthesized high-quality audio files, including 2,100 multi-track music pieces. Designed for audio source separation and multi-track audio modeling research, it provides rich multi-instrument training material for music information retrieval, audio separation, and music generation.

\textbf{•	MG-VAE (2020)}

The MG-VAE dataset\citep{luo2020mg}, developed by a research team from Xi'an Jiaotong University, includes over 2,000 MIDI-formatted Chinese folk songs representing both Han and minority regions. It employs Variational Autoencoder (VAE) methods to separate pitch and rhythm into distinct latent spaces of style and content, supporting music style transfer and cross-cultural music generation research.

\textbf{•	Groove2Groove (2020)}

The Groove2Groove dataset\citep{cifka2020groove2groove} is developed for music style transfer research, containing thousands of music audio clips with various styles and rhythms. It includes recordings of real instruments and synthesized audio, widely used in style transfer, music accompaniment generation, and automated arrangement studies.

\textbf{•	Hi-Fi Singer (2020)}

Developed by the HiFiSinger project team, this dataset focuses on high-fidelity singing voice synthesis research\citep{chen2020hifisinger}. It contains over 11 hours of high-quality singing recordings with a 48kHz sampling rate, addressing the challenges of high sampling rate modeling and fine acoustic details. It is widely used in high-quality singing voice synthesis, singing separation, and audio restoration research.

\textbf{•	MIDI-DDSP (2021)}

The MIDI-DDSP dataset\citep{wu2021midi} combines MIDI files and synthesized high-quality audio using Differentiable Digital Signal Processing (DDSP) technology. It is used in research on physically modeled music generation and synthesis, supporting applications in instrument modeling and audio generation requiring detailed control over musical expression.

\textbf{•	Singing Voice Conversion (2023)}

The Singing Voice Conversion dataset originates from the Singing Voice Conversion Challenge (SVCC 2023), derived from a subset of the NUS-HLT Speak-Sing dataset\citep{huang2023singing}. It includes singing and speech data from multiple singers, used for singing voice conversion and style transfer research, supporting the development of systems that can convert one singer's vocal style to another, essential for singing synthesis and imitation studies.

Please refer to Table \ref{table:3} for a comparison of the basic information of these datasets.
\begin{table*}[ht]
\centering
\begin{adjustbox}{max width=\textwidth}
\begin{tabular}{|p{3cm}|p{1.2cm}|p{2cm}|p{2.5cm}|p{4.5cm}|}
\hline
\textbf{Dataset Name} & \textbf{Year} & \textbf{Type} & \textbf{Scale} & \textbf{Main Application Areas} \\ 
\hline
CAL500 & 2007 & Audio & 500 songs & Emotion Recognition \\ 
\hline
MagnaTagATune & 2008 & Audio & 25,863 clips & Music Annotation, Emotion Recognition \\ 
\hline
Nottingham Music Dataset & 2009 & MIDI & 1000 tunes & Symbolic Music Analysis \\ 
\hline
Million Song Dataset & 2011 & Audio & 1,000,000 songs & Music Information Retrieval \\ 
\hline
MediaEval Emotion in Music & 2013 & Audio & 1000 songs & Emotion Recognition \\ 
\hline
AMG1608 & 2015 & Audio & 1608 clips & Emotion Recognition \\ 
\hline
VCTK Corpus & 2016 & Audio & 110 speakers & Speech Recognition, TTS \\ 
\hline
Lakh MIDI & 2017 & MIDI & 176,581 files & Music Information Retrieval \\ 
\hline
NSynth & 2017 & Audio & 300,000 samples & Music Synthesis \\ 
\hline
DEAM & 2017 & Audio & 1802 songs & Emotion Recognition \\ 
\hline
LJSpeech & 2017 & Audio & 13,100 clips & Speech Synthesis \\ 
\hline
Free Music Archive (FMA) & 2017 & Audio & 106,574 songs & Music Classification \\ 
\hline
AudioSet & 2017 & Audio & 2,000,000 clips & Audio Event Detection \\ 
\hline
CH818 & 2017 & Audio & 818 clips & Emotion Recognition \\ 
\hline
URMP & 2018 & Audio, Video, MIDI & 44 performances & Audio-Visual Analysis \\ 
\hline
MAESTRO & 2018 & MIDI, Audio & 200 hours & Music Generation, Piano Transcription \\ 
\hline
Groove MIDI Dataset & 2019 & MIDI, Audio & 13.6 hours & Rhythm Generation \\ 
\hline
GiantMIDI-Piano & 2020 & MIDI & 10,855 songs & Music Transcription, Analysis \\ 
\hline
LakhNES & 2019 & MIDI & 775,000 multi-instrument examples & Music Generation \\ 
\hline
Slakh2100 & 2019 & MIDI, Audio & 2100 tracks & Source Separation \\ 
\hline
MG-VAE & 2020 & MIDI & 2000 songs & Style Transfer \\ 
\hline
Groove2Groove & 2020 & Audio & thousands of clips & Style Transfer \\ 
\hline
Hi-Fi Singer & 2021 & Audio & 11 hours & Singing Voice Synthesis \\ 
\hline
MIDI-DDSP & 2022 & MIDI, Audio & varied & Music Generation, Synthesis \\ 
\hline
Singing Voice Conversion & 2023 & Audio & subset of NHSS & Voice Conversion \\ 
\hline
\end{tabular}
\end{adjustbox}
\caption{Overview of Music Datasets and Their Applications in AI Research}
\label{table:3}
\end{table*}

\textbf{5.2 Importance of Dataset Selection}

High-quality datasets not only provide rich training material but also significantly enhance the performance of generative models across different musical styles and complex structures. Therefore, careful consideration of the following key factors is essential when selecting and constructing datasets:

\textbf{•	Diversity:} A diverse dataset that covers a wide range of musical styles, structures, and expressions helps generative models learn different types of musical features. Diversity prevents models from overfitting to specific styles or structures, enhancing their creativity and adaptability in music generation. For example, the Lakh MIDI Dataset\citep{raffel2016learning} and NSynth Dataset\citep{engel2017neural} are popular among researchers due to their diversity, encompassing a broad repertoire from classical to pop music.

\textbf{•	Scale: }The scale of a dataset directly impacts a model’s generalization ability. Especially in deep learning models, large-scale datasets provide more training samples, enabling the model to better capture and learn complex musical patterns. This principle has been validated in many fields, such as Google Magenta’s use of large-scale datasets to train its generative models with significant results. For AI music generation, scale not only implies a large number of samples but also encompasses a broad range of musical styles and forms.

\textbf{•	Quality: }The quality of a dataset largely determines the effectiveness of music generation. High-quality datasets typically include professionally recorded and annotated music, providing accurate and high-fidelity training material for models. For example, datasets like MUSDB18\citep{stoter20182018} and DAMP (Digital Archive of Mobile Performances)\citep{smule2018damp} offer high-quality audio and detailed annotations, supporting precise training of music generation models.

\textbf{•	Label Information:} Rich label information (e.g., pitch, dynamics, instrument type, emotion tags) provides generative models with more precise contextual information, enhancing expressiveness and accuracy in generated music. Datasets with detailed labels, such as The GiantMIDI Dataset\citep{kong2020giantmidi}, include not only MIDI data but also detailed annotations of pitch, chords, and melody, allowing models to generate more expressive musical works.

\textbf{5.3 Challenges Faced by Datasets}
Despite their critical role in AI music generation, datasets face several challenges that limit current model performance and further research advancement:

\textbf{•	Dataset Availability: }High-quality and diverse music datasets are scarce, especially for tasks involving specific styles or high-fidelity audio generation. Publicly available datasets like the Lakh MIDI Dataset\citep{raffel2016learning}, while extensive, still lack data in certain specific music styles or high-fidelity audio domains. This scarcity limits model performance on specific tasks and hinders research progress in diverse music generation.

\textbf{•	Copyright Issues: }Copyright restrictions on music are a major barrier. Due to copyright protection, many high-quality music datasets cannot be publicly released, and researchers often have access only to limited datasets. This restriction not only limits data sources but also results in a lack of certain music styles in research. Copyright issues also affect the training and evaluation of music generation models, making it challenging to generalize research findings to broader musical domains.

\textbf{•	Dataset Bias: }Music styles and structures within datasets often have biases, which can result in generative models producing less diverse outputs or favoring certain styles. For example, if a dataset is dominated by pop music, the model may be biased toward generating pop-style music, overlooking other types of music. This bias not only affects the model's generalization ability but also limits its performance in diverse music generation.

\textbf{5.4 Future Dataset Needs}
With the development of AI music generation technologies, the demand for larger, higher-quality, and more diverse datasets continues to grow. To drive progress in this field, future dataset development should focus on the following directions:

\textbf{•	Multimodal Datasets: }Future research will increasingly focus on the use of multimodal data. Datasets containing audio, MIDI, lyrics, video, and other modalities will provide critical support for research on multimodal generative models. For example, the AudioSet Dataset\citep{gemmeke2017audio}, as a multimodal audio dataset, has already demonstrated potential in multimodal learning. By integrating various data forms, researchers can develop more complex and precise generative models, enhancing the expressiveness of music generation.

\textbf{•	Domain-Specific Datasets: }As AI music generation technology becomes more prevalent across different application scenarios, developing datasets targeted at specific music styles or applications is increasingly important. For instance, datasets focused on therapeutic music or game music will aid in advancing research on specific tasks within these fields. The DAMP Dataset\citep{smule2018damp}, which focuses on recordings from mobile devices, provides a foundation for developing domain-specific music generation models.

\textbf{•	Open Datasets: }Encouraging more music copyright holders and research institutions to release high-quality datasets will be crucial for driving innovation and development in AI music generation. Open datasets not only increase data availability but also foster collaboration among researchers, accelerating technological advancement. Projects like Common Voice\citep{ardila2019common} and Freesound\citep{fonseca2017freesound} have significantly promoted research in speech and sound recognition through open data policies. Similar approaches in the music domain will undoubtedly lead to more innovative outcomes.

By making progress in these areas, the AI music generation field will gain access to richer and more representative data resources, driving continuous improvements in music generation technology. These datasets will not only support more efficient and innovative model development but also open up new possibilities for the practical application of AI in music creation.

\section{Evaluation Methods}
Evaluating the quality of AI-generated music has always been a focus of researchers. Since the early days of computer-generated music, assessing the quality of these works has been a key issue. Below are the significant research achievements at different stages.

\textbf{6.1 Overview of Evaluation Methods}

In terms of subjective evaluation, early research relied heavily on auditory judgments by human experts, a tradition dating back to the 1970s to 1990s. For example, \citep{loy1985programming}evaluated computer-generated music clips through listening tests. By the 2000s, subjective evaluation methods became more systematic. \citep{cuthbert2010music21} proposed a survey-based evaluation framework to study the emotional and aesthetic values of AI-generated music. With the advancement of deep learning technologies, the complexity of subjective evaluation further increased. \citep{papadopoulos2016assisted} and \citep{yang2017midinet} introduced multidimensional emotional rating systems and evaluation models combining user experience, marking a milestone in subjective evaluation research. Recently, \citep{agarwal2021efficient} proposed a multi-level evaluation framework based on emotion recognition, and \citep{chu2022empirical} developed a user satisfaction measurement tool, which more accurately captures complex emotional responses and cultural relevance, making subjective evaluation methods more systematic and detailed.

Objective evaluation dates back to the 1980s when the quality of computer-generated music was assessed mainly through a combination of audio analysis and music theory. Cope \citep{cope1996experiments}pioneered the use of music theory rules for structured evaluation. Subsequently, Huron \citep{huron2008sweet} introduced a statistical analysis-based model for evaluating musical complexity and innovation, quantifying structural and harmonic features of music, thus providing important tools for objective evaluation. With the advent of machine learning, Conklin \citep{conklin2003music} and Briot et al. \citep{briot2017deep} developed more sophisticated objective evaluation systems using probabilistic models and deep learning techniques to analyze musical innovation and emotional expression.

\textbf{6.2 Evaluation of Originality and Emotional Expression}

The evaluation of originality became an important research direction in the 1990s. \citep{miranda1995artificial} and \citep{toiviainen2006autocorrelation} introduced early mechanisms for originality scoring through genetic algorithms and computational models. As AI technology advanced, researchers such as \citep{herremans2017functional} combined Markov chains and style transfer techniques, further enhancing the systematic and diverse evaluation of originality.
The evaluation of emotional expression began with audio signal processing. \citep{sloboda1991music} and \citep{picard2000affective} laid the foundation for assessing emotional expression in music through the analysis of pitch, rhythm, and physiological signals. With the rise of multimodal analysis, \citep{kim2010music} and \citep{yang2012machine} developed emotion analysis models that combine audio and visual signals, significantly improving the accuracy and diversity of emotional expression evaluation.

\textbf{6.3 Implementation Strategies of Evaluation Frameworks}

The implementation strategies of evaluation frameworks have evolved from simple to complex. The combined use of qualitative and quantitative analysis was first proposed by Reimer\citep{reimer1991philosophy} in the field of music education and later widely applied in the evaluation of AI-generated music. Modern evaluation frameworks, such as those by Lim et al. (2017), integrate statistical analysis with user feedback, offering new approaches for comprehensive evaluation of AI-generated music. Multidimensional rating systems originated from automated scoring in films and video content, with \citep{hastie2009elements} laying the groundwork for multidimensional rating models in music evaluation. \citep{herremans2016tension} further extended this concept to the evaluation of music creation quality. Interdisciplinary collaboration and customized evaluation tools have become increasingly important in recent AI music evaluation. Research by \citep{gabrielsson2001emotion} emphasized the significance of cross-disciplinary collaboration in developing evaluation tools tailored to different styles and cultures. Finally, automated evaluation and real-time feedback, as key directions in modern music evaluation, have significantly enhanced the efficiency and accuracy of music generation quality assessment through machine learning and real-time analysis technologies.

\textbf{6.4 Conclusion}

By integrating subjective and objective evaluation methods and considering originality and emotional expressiveness, a comprehensive quality evaluation framework can be constructed. The early research laid the foundation for current evaluation methods, and recent advancements, particularly in evaluating originality and emotional expression, have achieved notable success. This comprehensive evaluation approach helps to more accurately measure the performance of AI music generation systems and provides guidance for future research and development, advancing AI music generation technology toward the complexity and richness of human music creation.

\section{Application Areas}

AI music generation technology has broad and diverse applications, from healthcare to the creative industries, gradually permeating various sectors and demonstrating immense potential. Based on its development history, the following is a detailed description of various application areas and the historical development of relevant research.

\textbf{7.1 Healthcare}

AI music generation technology has gained widespread attention in healthcare, particularly in emotional regulation and rehabilitation therapy. In the 1990s, music therapy was widely used to alleviate stress and anxiety. \citep{standley1986music} studied the effect of music on anxiety symptoms and highlighted the potential of music as a non-pharmacological treatment method. Although the focus was mainly on natural music at the time, \citep{sacks2008musicophilia}, in his book Musicophilia, further explored the impact of music on the nervous system, indirectly pointing to the potential of customized music in neurological rehabilitation. With advancements in AI technology, generated music began to be applied in specific therapeutic scenarios. \citep{aalbers2017music} demonstrated the positive impact of music therapy on emotional regulation and proposed personalized therapy through AI-generated music.

\textbf{7.2 Content Creation}

Content creation is one of the earliest fields where AI music generation technology was applied, evolving from experimental uses to mainstream creative tools. In the 1990s, David Cope’s Experiments\citep{cope1996experiments} in Musical Intelligence (EMI) (1996) was an early attempt at using AI-generated music for content creation. EMI could simulate various compositional styles, and its generated music was used in experimental works. Although the technology was still relatively basic, this pioneering research laid the foundation for future applications. In the 2000s, AI-generated music began to be widely used in creative industries like film and advertising. Startups such as Jukedeck developed music generation platforms using Generative Adversarial Networks (GANs) and Recurrent Neural Networks (RNNs) to create customized background music for short videos and ads. Briot et al.  found that AI-generated music had approached human-created music in quality and complexity, highlighting AI’s potential to improve content creation efficiency\citep{briot2020deep}. Recently, AI music generation technology has been applied even more widely in content creation. OpenAI’s MuseNet\citep{payne2019musenet} and Google’s Magenta project\citep{magenta2023} demonstrated the ability to generate complex, multi-style music, providing highly context-appropriate background music for films, games, and advertisements.

\textbf{7.3 Education}

AI music generation technology has revolutionized music education, becoming an important tool for understanding music theory and practical composition. In the early 21st century, AI began to be applied in music education. Pachet  explored the potential of automatic composition software in education, generating simple exercises to help students understand music structures and harmonies\citep{pachet2003continuator}. These early systems aimed to assist rather than replace traditional teaching methods. As technology advanced, AI music generation systems became more intelligent and interactive. Platforms such as MusEDLab’s AI Duet and Soundtrap’s AI Music Tutor\citep{musedlab} provide interactive educational experiences, listening to users’ performances, interpreting inputs, and offering instant feedback or real-time performance to help improve skills and understand musical nuances.

\textbf{7.4 Social Media and Personalized Content}

AI-generated music significantly enriches user experiences in social media and personalized content, with personalized recommendations and automated content generation becoming key trends. In the 2000s, social platforms like MySpace first introduced simple music generation algorithms to create background music for user profiles. Although technically basic, these early attempts laid the groundwork for personalized content generation. As social media platforms diversified, personalized content generation became mainstream. Music streaming platforms like Spotify and Pandora use AI to generate personalized playlists by analyzing user listening history and preferences, providing highly customized music experiences. AI-generated music is also used on short video platforms to enhance content appeal. Recently, AI-generated music has become an essential part of social media, with platforms like TikTok using AI to generate background music that quickly matches video content, significantly enhancing user experience. The personalized capabilities of AI-generated music greatly enhance user engagement and interaction on social media\citep{singhmedia}. Furthermore, its applications in virtual reality (VR) and augmented reality (AR) elevate immersive experiences, offering users novel sensory enjoyment.

\textbf{7.5 Gaming and Interactive Entertainment}

In gaming and interactive entertainment, AI music generation technology not only improves music creation efficiency but also enhances player immersion. Game developers began exploring algorithmic background music generation in the 1990s. For instance, The Sims series used procedural music generation that dynamically adjusted background music based on player actions and emotional states, laying the foundation for later game music generation. As games became more complex, AI music generation found broader applications in gaming. The concept of procedural audio was introduced into games, with Björk et al.  exploring music generation in interactive environments\citep{bjork2005patterns}. By the 2010s, AI technology had evolved to enable dynamic music generation that could adapt in real-time to game environments and player interactions, particularly in open-world and massively multiplayer online games (MMORPGs). Recent studies, such as those by Foley et al. (2023), highlight AI-generated music’s role in dynamically creating appropriate background music based on player behavior and emotions, enhancing player immersion and interaction. AI-generated music and sound effects in games not only improve the gaming experience but also reduce development time and costs\citep{beatoven2023}.

\textbf{7.6 Creative Arts and Cultural Industries}

AI-generated music has shown unique potential in the creative arts and cultural industries, pushing the boundaries of artistic creation. Xenakis  combined algorithms with music composition\citep{xenakis1992formalized}, ushering in a new era of computer-assisted creativity, providing theoretical foundations and practical experience for AI’s application in the arts. Briot et al.  discussed AI’s potential in generating complex musical forms\citep{briot2020deep}, applied in modern art and experimental music creation, showcasing AI-generated music’s broad applications in creative arts. Recently, AI-generated music has reached new heights in creative arts. Modern artists use AI technology to produce experimental music, breaking traditional boundaries of music composition. AI-generated music is also applied in dance choreography and theater scoring, enhancing the expressiveness of performing arts. In NFT (Non-Fungible Token) artworks, AI-generated music is part of the creation and sales process, driving new forms of digital art.

\textbf{7.7 Broadcasting and Streaming}

The application of AI-generated music in broadcasting and streaming services is expanding, significantly enhancing content richness and personalization. Early streaming platforms like Pandora and Last.fm used simple algorithms to generate recommended playlists based on user listening history, laying the foundation for later AI-generated music in streaming. By the 2010s, streaming services like Spotify began using deep learning and machine learning technologies to generate personalized music recommendations. Spotify’s Discover Weekly feature, a prime example, combines AI-generated music with recommendation systems to deliver highly customized music experiences. Recently, AI-generated music’s application in broadcasting and streaming has become more complex and diverse. For instance, AI-generated background music is used in news broadcasts and podcasts, enhancing the emotional expression of content. Streaming platforms also use AI-generated music to create seamless playlists tailored to different user contexts, such as fitness, relaxation, or work settings. AI-generated new music styles and experimental music offer users unprecedented auditory experiences.

\textbf{7.8 Marketing and Brand Building}

AI-generated music has unique applications in marketing and brand building, enhancing brand impact through customized music. In early brand marketing, background music was typically chosen by human planners, but with the development of AI technology, companies began exploring AI-generated music to enhance advertising impact. Initial applications focused on generating background music for ads to increase brand appeal. By the 2010s, AI-generated music became more common in advertising. Startups like Amper Music developed AI music generation platforms that help companies generate music aligned with their brand identity, strengthening emotional connections with audiences. Recently, the application of AI-generated music in brand building has deepened. Brands can use AI-generated music to create unique audio identities, enhancing brand recognition. AI-generated music is also widely used in cross-media marketing campaigns, seamlessly integrating with video, images, and text content, offering new ways to tell brand stories. Moreover, AI-generated music is used in interactive ads to create real-time background music that interacts with consumers, further strengthening brand-consumer connections.

AI music generation technology has shown significant value across multiple fields. From healthcare to content creation, education to social media, AI not only improves music generation efficiency but also greatly expands the scope of music applications. As technology continues to advance, AI music generation will play an increasingly important role in more fields, driving comprehensive innovation in music creation and application. These applications demonstrate AI’s innovative potential in music generation and highlight its importance in improving human quality of life, enhancing creative efficiency, and promoting cultural innovation.

\section{Challenges and Future Directions}
Despite significant progress in AI music generation technology, multiple challenges remain, providing rich avenues for future exploration. The current technological bottlenecks are primarily centered on the following key issues:

Firstly, the diversity and originality of generated music remain major concerns for researchers. Early generative systems, such as David Cope’s Experiments in Musical Intelligence (EMI)\citep{cope2023emi}, were successful at mimicking existing styles but often produced music that was stylistically similar and lacked innovation. This limitation in diversity has persisted in later deep learning models. Although the introduction of Generative Adversarial Networks (GANs) and Recurrent Neural Networks (RNNs) improved diversity, the results still often suffer from “mode collapse”—where generated pieces across samples are too similar in style, lacking true originality. This phenomenon was extensively discussed in Briot et al., highlighting the potential limitations of deep learning models in music creation\citep{briot2020deep}.

Secondly, effectively capturing long-term dependencies and complex structures in music is a critical challenge in AI music generation\citep{briot2020deep}. As a time-based art form, music’s structure and emotional expression often rely on complex temporal spans and hierarchies\citep{hawthorne2018enabling}. Current AI models struggle with this complexity, and although some studies have attempted to address this by increasing the number of layers in the model or introducing new architectures (such as Transformer models), results show that models still find it difficult to generate music with deep structural coherence and long-term dependencies. The core issue is how to enable models to maintain overall macro coherence while showcasing rich details and diversity at the micro level during music generation.

The standardization of evaluation methods has also been a persistent challenge in assessing the quality of AI-generated music. Traditional evaluation methods mainly rely on subjective assessments by human listeners, but these methods often lack consistency and objectivity\citep{yang2012machine}. With the expanding applications of AI-generated music, the need for more objective and consistent evaluation standards has grown. Researchers have begun exploring quantitative evaluation methods based on statistical analysis and music theory\citep{herremans2016tension} however, effectively integrating these methods with subjective assessments remains an area needing further exploration\citep{engel2017neural}. The refinement of such evaluation systems is crucial for advancing the practical applications of AI music generation technology.

Facing these challenges, future research directions can focus on the following areas:

\textbf{Exploring New Music Representations and Generation Methods:} Introducing more flexible and diverse music representation forms can help generative models better capture the complexity and diversity of music. Research in this area can draw on recent findings in cognitive science and music theory to develop generation mechanisms that better reflect the human creative process.

\textbf{Enhancing Control Capabilities of Hybrid Models:} By incorporating more contextual information (such as emotion tags or style markers), AI-generated music can achieve greater progress in personalization and diversity. The control capabilities of hybrid models directly affect the expressiveness and user experience of generated music, making this a critical direction for future research.

\textbf{Applying Interdisciplinary Approaches: }Combining music theory, cognitive science, and deep learning will be key to advancing AI music generation. This approach can enhance the ability of generative models to capture complex musical structures and make AI-generated music more aligned with human aesthetic and emotional needs. Interdisciplinary collaboration can lead to the development of more intelligent and human-centered music generation systems.

\textbf{Real-Time Generation and Interaction:} Real-time generation and adjustment of music will bring unprecedented flexibility and creative space to music creation and performance. Particularly in interactive entertainment and live performances, real-time generation technology will significantly enhance user experience and artistic expressiveness.

By conducting in-depth research in these directions, AI music generation technology is expected to overcome existing limitations, achieving higher levels of structural coherence, expressiveness, and diversity, thus opening new possibilities for music creation and application. This will not only drive the intelligent evolution of music creation but also profoundly impact the development of human music culture.

\section{Conclusion}
This paper provides a comprehensive review of the key technologies, models, datasets, evaluation methods, and application scenarios in the field of AI music generation, offering a series of summaries and future directions based on the latest research findings. By reviewing and analyzing existing studies, this paper presents a new summarization framework that systematically categorizes and compares different technological approaches, including symbolic generation, audio generation, and hybrid models, thereby offering researchers a clear overview of the field. Through extensive research and analysis, this paper covers emerging topics such as multimodal datasets and emotional expression evaluation and reveals the potential impact of AI music generation across various application areas, including healthcare, education, and entertainment.

However, despite significant advances in the diversity, originality, and standardization of evaluation methods, AI music generation technology still faces numerous challenges. In particular, capturing complex musical structures, handling long-term dependencies, and ensuring the innovation of generated music remain pressing issues. Future research should focus more on the diversity and quality of datasets, explore new generation methods, and promote interdisciplinary collaboration to overcome the current limitations of the technology.

Overall, this paper provides a comprehensive knowledge framework for the field of AI music generation through systematic summaries and analyses, offering valuable references for future research directions and priorities. This not only contributes to the advancement of AI music generation technology but also lays the foundation for the intelligent and diverse development of music creation. As technology continues to evolve, the application prospects of AI in the music domain will become even broader. Future researchers can build upon this work to further expand the field, bringing more innovation and breakthroughs to music generation.

\bibliography{arxiv}

\end{document}